\documentclass{scrartcl}
\usepackage[utf8]{inputenc}
\usepackage{enumitem}
\usepackage{hyperref}
\usepackage{caption}
\usepackage{subcaption}
\usepackage{amssymb}
\usepackage{graphicx}
\usepackage{url}
\usepackage{array}
\newcolumntype{P}[1]{>{\raggedleft\arraybackslash}p{#1}}
\usepackage{footnote}
\usepackage{bm}
\usepackage{multirow}
\usepackage{tabularx}
\usepackage{booktabs}
\usepackage{color}
\usepackage{amsmath}

\newcommand{\ignore}[1]{}

\newcommand{\fun}{\mathit{fun}}

\newcommand{\complete}{\mathit{complete}}
\newcommand{\incomplete}{\mathit{incomplete}}

\newcommand{\precision}{\mathit{precision}}
\newcommand{\recall}{\mathit{recall}}
\newcommand{\true}{\mathit{true}}
\newcommand{\cwa}{\mathit{cwa}}
\newcommand{\pca}{\mathit{pca}}
\newcommand{\pop}{\mathit{pop}}
\newcommand{\chg}{\mathit{chg}}
\newcommand{\moreThan}{\mathit{moreThan}}
\newcommand{\lessThan}{\mathit{lessThan}}
\newcommand{\notype}{\mathit{notype}}
\newcommand{\hasChild}{\mathit{hasChild}}
\newcommand{\type}{\mathit{type}}
\newcommand{\popularity}{\mathit{popularity}}
\newcommand{\nochange}{\mathit{nochange}}
\newcommand{\card}{\mathit{card}}
\renewcommand{\star}{\mathit{star}}
\newcommand{\class}{\mathit{class}}
\makeatletter
\newcommand*{\defeq}{\mathrel{\rlap{%
  \raisebox{0.3ex}{$\m@th\cdot$}}%
  \raisebox{-0.3ex}{$\m@th\cdot$}}%
  =}
\makeatother

\makeatletter
\DeclareOldFontCommand{\rm}{\normalfont\rmfamily}{\mathrm}
\DeclareOldFontCommand{\sf}{\normalfont\sffamily}{\mathsf}
\DeclareOldFontCommand{\tt}{\normalfont\ttfamily}{\mathtt}
\DeclareOldFontCommand{\bf}{\normalfont\bfseries}{\mathbf}
\DeclareOldFontCommand{\it}{\normalfont\itshape}{\mathit}
\DeclareOldFontCommand{\sl}{\normalfont\slshape}{\@nomath\sl}
\DeclareOldFontCommand{\sc}{\normalfont\scshape}{\@nomath\sc}
\makeatother

\begin{document}
\title{Predicting Completeness in\\Knowledge Bases}

\author{
\begin{tabular}[t]{c}
  Luis Gal\'arraga\\ 
{\normalfont Aalborg University, Denmark} \\[0.5em]
Simon Razniewski\\
{\normalfont }Free University of Bozen-Bolzano, Italy \\[0.5em]
Antoine Amarilli \\
  {\normalfont Télécom ParisTech, Université Paris-Saclay, France}\\[0.5em]
Fabian M.\ Suchanek \\
  {\normalfont Télécom ParisTech, Université Paris-Saclay, France}\\[0.5em]
\end{tabular}
}
\date{}

\maketitle

\begin{abstract}
  Knowledge bases such as Wikidata, DBpedia, or YAGO contain millions of entities and facts. In some
  knowledge bases, the
  \emph{correctness} of these facts has been evaluated. However, much less is known about
  their \emph{completeness}, i.e., the proportion of real facts that the
  knowledge bases cover.
  In this work, we investigate different signals to identify the areas where a knowledge base is complete.
  We show that we can combine these signals
  in a rule mining approach, which allows us to predict 
  where facts may be missing. We also show that completeness
  predictions can help other applications
  such as fact prediction.
\end{abstract}

\pagebreak

\section{Introduction}
\label{sec:introduction}
\paragraph{Motivation} Knowledge Bases (KBs) such as DBpedia~\cite{dbpedia}, NELL~\cite{nell}, Wikidata~\cite{wikidata}, the Google Knowledge Vault~\cite{knowledge-vault}, or YAGO~\cite{yago} contain billions of machine-readable facts about the world.
They know for instance that Paris is the capital of France and that Barack Obama won the Nobel Peace Prize.
KBs have applications in information retrieval, question answering, machine translation, and data maintenance, among others.

However, the data quality of KBs is not always perfect. Problems include false
data, missing information, or schema inconsistencies. Hence, many
approaches aim to clean up erroneous information~\cite{quality-assessment}.
In contrast, the completeness (recall) of the KBs has remained relatively unexplored. While we often know what proportion of the facts in the KB are correct, we usually do not know what proportion of the facts in the real world they cover.

For example, as of 2016, Wikidata knows the father of only 2\% of all people in the KB~-- even though in the real world everyone has a father. DBpedia contains only 6 Dijkstra Prize winners -- but in the real world there are 35. Likewise, according to YAGO, the average number of children per person is 0.02. In general, between 69\% and 99\% of instances in popular KBs lack at least  one  property  that  other  entities  in  the  same class have \cite{watermarking,distant-supervision}. Thus, we know that today's KBs are highly incomplete, but we do not know where the information is missing.

This unknown degree of completeness poses several problems~\cite{vision-paper}. First, users do not have any guarantee that a query run against the KB yields all the results that match the query in the real world.
Second, the data providers themselves may not know where the data is incomplete,
and thus cannot determine where to focus their efforts. If  they  knew, e.g., which people are missing their alma mater, they could focus on tracing these pieces of information and adding them. Third, completeness information could help identify wrong facts. If we knew, e.g., 
that people always have only 2 parents,
then a KB that contains 
3 parents for an individual has to be erroneous. Finally, completeness information can be insightful on its own, to know which missing facts are known to be wrong. This is useful, e.g., for machine learning algorithms that require counter-examples.

Thus, it would be of tremendous use for both data providers and data consumers if we could know where the information in the KB is complete. In the ideal case, we would want to make what we call \emph{completeness assertions}, which say, e.g., \emph{This KB contains all children of Barack Obama}.

\paragraph{Challenges} The main obstacle to establish such completeness assertions is the Open World Assumption (OWA), which nearly all KBs make.
The OWA says that if the KB does not contain a certain piece of information, then this information is not necessarily false -- it may be true in the real world, but absent from the KB.
This means that every part of the KB could be potentially incomplete.
Furthermore, today's KBs mostly consist of subject-predicate-object triples.
These formalisms usually provide very limited means to store negative information (if at all).
For example, YAGO says that Barack Obama is married to Michelle Obama, but it does not say that Barack Obama is not (and was never) married to any other person. In fact, there is not even a way that YAGO and similar KBs could express this idea.
The KBs are not just incomplete, but also, by design, unable to provide any indications of completeness.

\paragraph{Contribution} In this paper, we make a first step towards generating completeness information \emph{automatically}. Our goal is to determine automatically whether certain properties of certain objects are complete: whether a person has more children in reality than in the KB, whether a person graduated from a university in real life even though the KB does not know about it, or whether a person has more spouses in reality than are known to the KB.
More precisely:
\begin{itemize}
\item We conduct a systematic study of signals that can indicate completeness of properties of objects in a KB.
\item We show how completeness assertions can be learned through a rule mining
  system, AMIE; we further show how the necessary training data for AMIE can be
    obtained easily through
    crowdsourcing.
\item We find that completeness can be predicted for some relations with up to 100\% precision on real KBs (YAGO and Wikidata).
\item As a use case, we show that our completeness assertions can increase the precision of rule mining.
\end{itemize}
This paper is structured as follows. We first discuss related work in Section~\ref{sec:relatedwork}, and introduce
preliminaries in Section~\ref{sec:preliminaries}. We then present in
Section~\ref{sec:estimatingcompleteness} the different signals that we use
to predict completeness, and leverage them in
Section~\ref{sec:learningcompleteness} to mine completeness rules with the AMIE system. Section~\ref{sec:experiments} presents detailed evaluations of the signals in isolation and in combination. We showcase in Section~\ref{sec:usingcompleteness} an application of completeness assertions, before concluding in Section~\ref{sec:conclusion}.

\section{Related Work}
\label{sec:relatedwork}
\paragraph{Knowledge Bases}
Many of today's KBs provide estimations of their precision. The YAGO KB~\cite{yago} was manually evaluated and found to be 95\% correct. NELL~\cite{nell} is regularly checked by humans for precision. Facts in the Knowledge Vault~\cite{knowledge-vault} are annotated with an estimated precision.
However, little is known about the recall/completeness of these KBs. Of course, larger sizes may indicate higher completeness, but size is only a very coarse proxy for completeness.

\paragraph{Incompleteness Studies} Some studies have found that KBs are indeed quite incomplete. For instance, a watermarking study~\cite{watermarking} reports that 69\%--99\% of instances in popular KBs lack at least one property that other entities in the same class have. Google found that  71\%  of  people  in Freebase  have  no  known  place  of  birth,  and  75\%  have no known nationality~\cite{knowledge-vault}. This allows us to know that KBs are incomplete in general, but not which parts are complete.

\paragraph{Manual Indicators} The Wikidata community maintains lists that explain where information is still missing -- e.g., a list of people without birth dates\footnote{\scriptsize \url{https://www.wikidata.org/wiki/Wikidata:Database_reports/top_missing_properties_by_number_of_sitelinks/P569}}.
Also, Wikidata contains \emph{no-value statements}, which say that an empty relation is complete for an entity~\cite{novalues}. An extension for Wikidata allows contributors to manually add recall information~\cite{cool-wd}. However, these annotations are mostly provided manually: our work aims at deducing 
such annotations \emph{automatically}.

\paragraph{Partial Completeness Assumption} Some approaches simply \emph{assume} that KBs are complete in certain areas. For instance, the AMIE project used the \emph{partial completeness assumption} (PCA)~\cite{amie-plus} (re-used as the \emph{local closed world assumption} in \cite{knowledge-vault}). We discuss the PCA in detail in Section~\ref{sec:estimatingcompleteness}.

\paragraph{Rule Mining} Inductive Logic Programming and Rule Mining approaches~\cite{ilp} find rules such as \emph{If a person lives in a city, then their spouse probably lives in the same city}. These rules can then be used to predict new information (here: where the spouse lives). As a side effect, this procedure determines where the KB is incomplete. However, such approaches can only ever mine new facts between instances that are already known to the KB. They cannot tell us that a spouse is missing if that spouse is not in the KB. We will show in our experiments how rule mining approaches can benefit from the techniques we develop in this paper.

\paragraph{Completeness Reasoning} On the database side, some work has investigated how to combine completeness information about parts of databases to deduce completeness annotations on query results~\cite{motro_integrity,levy,razniewskietal:sigmod2015}. However, this work assumes that the KB has already been annotated with completeness assertions. Our goal, in contrast, is to \emph{generate} such assertions.

\section{Preliminaries}
\label{sec:preliminaries}
\paragraph{Knowledge Bases} In this paper, we target KBs in RDFS format~\cite{rdfs}. We
assume that the reader is familiar with RDFS. We write facts as $r(s, o)$, where $r$ is a relation, $s$ is the subject, and $o$ is the object. For instance, $\mathit{president}(\mathit{Obama}, \mathit{USA})$ is a fact.
We assume a fixed KB~$\mathcal{K}$, and thus write $r(s, o)$ to mean $r(s, o) \in \mathcal{K}$.

\paragraph{Functionality} The \emph{functionality}~\cite{paris} of a relation $r$ is defined as:
\[\fun(r) \defeq \frac{\#x: \exists y: r(x, y)}{\#(x, y): r(x, y)}\]
where $\#\alpha: \mathcal{A}$ denotes the number of $\alpha$ that fulfill the condition $\mathcal{A}$. For relations such as \emph{place\-Of\-Birth} which are functions, we have $\fun(r)=1$. For ``quasi-functions'' such as \emph{is\-Citizen\-Of}, the value $\fun(r)$ is close to 1. If $r$ has many objects for a subject, then $\fun(r)$ is closer to 0.

\paragraph{Completeness} \label{subsubsec:completeness} In line with work in databases~\cite{motro_integrity,vision-paper}, we define completeness via a hypothetical ideal KB $\mathcal{K}^*$, which contains all facts of the real world. A KB $\mathcal{K}$ is \emph{complete} for a query $q$, if $q$ returns the same results on $\mathcal{K}$ as on $\mathcal{K}^*$. In this paper, we focus on a particular type of queries, namely those that ask for the objects of a given subject and relation. Thus, a pair of an entity $s$ and a relation $r$ is \emph{complete} in a KB $\mathcal{K}$, if $\{o : r(s, o) \in \mathcal{K}\} \supseteq \{o : r(s, o) \in \mathcal{K}^*\}$. For example, a KB is complete for the subject \emph{Barack Obama} and the relation \emph{has\-Child}, if it contains both of Obama's children. If the KB is complete for a subject $s$ and a relation $r$, we make a \emph{completeness assertion} of the form $\complete(s, r)$. Our goal is to establish such completeness assertions.

In general, completeness assertions make less sense for relations with low functionality. For example, it does not make sense to ask a KB if it knows all citizens of France. It is more sensible to ask whether the KB knows all nationalities of one person. Therefore, we consider completeness primarily for relations with high functionality. In particular, if a relation has low functionality (such as \emph{country\-Has\-Citizen}), and its inverse has high functionality (\emph{person\-Has\-Nationality}), then we consider the inverse.

When a relation is incomplete for a subject, we could also try to estimate \emph{how many} objects are missing. This would amount to a cardinality estimation. In this paper, however, we focus on the simpler task of establishing completeness assertions, and leave cardinality estimations for future work.

\paragraph{Completeness Considerations} The notion of completeness is not well-defined for all relations~\cite{vision-paper}. Take, e.g., the relation \emph{hasHobby}. It is not always clear whether an activity counts as a hobby or not.  Thus, it is difficult to establish whether a KB is complete on the hobbies of a certain person. Even for seemingly well-defined relations such as \emph{has\-Official\-Language}, completeness is not easy to establish: a country may have de facto official languages that are not legally recognized (e.g., the US); languages that are official in some regions but not in the country (e.g., India); or an official language that is not a spoken language (e.g., New Zealand). In this paper, we manually selected relations for which completeness is well-defined, and concentrate on these.

\newcommand{\compl}{\mathit{compl}}

\paragraph{Completeness Oracles}
\label{subsubsec:completeness-oracles}
A \emph{completeness oracle} tries to guess whether a given relation is complete for a given subject in the fixed KB $\mathcal{K}$. Technically, a completeness oracle is a binary relation on entities and relations that holds whenever the oracle predicts that a given subject is complete for a given relation.
The Partial Completeness Assumption (PCA) is an example of a simple completeness oracle. 
It predicts completeness for a subject $s$ and a relation~$r$ if there exists an object $x$ with $r(s, x)$. 
For instance, if in a KB, Barack Obama has one child, the PCA oracle will (wrongly) state that Barack Obama 
is complete for the relation \emph{has\-Child}, i.e., \emph{pca(Barack\-Obama, has\-Child)} will be true.

The precision and recall of an oracle $o$ are defined as follows, where \emph{complete} denotes the completeness assertions on~$\mathcal{K}$ that are true relative to the ideal KB $\mathcal{K}^*$:
\begin{align*}
  \precision(o) & \defeq \frac{\#(e, r): o(e, r) \wedge \complete(e, r)}{\#(e, r): o(e, r)}\\[.5em]
  \recall(o) & \defeq \frac{\#(e, r): o(e, r) \wedge \complete(e, r)}{\#(e, r): \complete(e, r)}
\end{align*}
\noindent The \emph{F1 measure} is defined as usual from precision and recall.

\section{Completeness Oracles}
\label{sec:estimatingcompleteness}
We now present various completeness oracles, of which we study two
kinds: \emph{simple} oracles and \emph{parameterized} oracles.

\subsection{Simple Oracles}
\label{subsec:simpleOracles}
\paragraph{Closed World Assumption} The Closed World Assumption (CWA) assumes that any fact that is not in the KB does not hold in the real world. That is, the CWA assumes that the entire KB is complete. In general, the CWA is incompatible with the philosophy of the Semantic Web. Still, the CWA may be suitable under certain conditions. For instance, if a person is not known to be the president of any country, then most likely the person is indeed not the president of any country.
Formally, the CWA completeness oracle is simply defined as:
\[\cwa(s, r)\defeq \true\]

\paragraph{Partial Closed World Assumption (PCA)}
The PCA~\cite{amie} is an oracle that has proven useful for rule mining~\cite{knowledge-vault,amie-plus}.
Under the PCA, a subject-relation pair $s, r$ is considered complete if there is at least an object $o$ with $r(s, o)$. In other words, we assume that, if the KB knows some $r$-values for $s$, then it knows all its values. The PCA is more cautious at predicting completeness than the CWA: it predicts completeness only if objects are already known. This implies that the PCA makes predictions only for those entities that have an object for the relationship, and remains silent otherwise.
For instance, according to the CWA, a person that has no nationality in the KB has no nationality in reality, but the PCA will not make such claims.
Formally, the \emph{PCA oracle} is:
\[\pca(s, r)\defeq \exists o: r(s, o)\]
\noindent The PCA is especially well suited for \emph{functional relations}, where an entity can have at most one object. Indeed, if an entity has some object for a functional relation, then it is necessarily complete.

\paragraph{Cardinality}
A more cautious oracle than the PCA is the \emph{cardinality oracle}. For
an integer value $k$, the cardinality oracle for value~$k$ says that a subject $s$ is complete for a relation~$r$ if $s$ has at least $k$ different objects for~$r$. Formally:
\[\card_k(s, r)\defeq \#(o: r(s, o)) \geq k\]
\noindent This oracle subsumes the CWA and PCA: $\card_0$ is $\cwa$, and $\card_1$ is $\pca$. Other values for $k$ can be useful, e.g., $\card_2$ can be effectively used as a predictor for the $\mathit{hasParent}$ relation. In our experience, however, larger values of~$k$ are rarely useful, and hence we categorize this oracle as a simple oracle.

\paragraph{Popularity} The previous oracles look at properties of entities in isolation, but
we can also look at entities in the context of the entire KB. For example, we can hypothesize that entities which are popular (by some measure) are more likely to be complete.
For example, we expect that Wikipedia-based KBs are more complete for famous
entities (e.g., Albert Einstein) than for entities that have only stub-articles.
From a Boolean measure $\pop$ indicating whether an
entity is popular or not, we define the \emph{popularity oracle}~as:
\[\popularity_{\pop}(s, r) \defeq \pop(s)\]

\paragraph{No Change} So far, we have only looked at a single snapshot of the
KB, but we can also study how the KB changes over time.
If the objects of a particular subject do not change, then this may suggest that the subject is complete. Given a Boolean
measure of change $\chg$, where $\chg(s, r)$ indicates whether the set of objects for entity $s$ and relation~$r$ has changed over time, we define the \emph{no-change oracle} by:
\[\nochange_{\chg}(s, r)\defeq\neg \chg(s, r)\]

\subsection{Parameterized Oracles}

We now move on to the study of oracles that depend on parameters that are difficult to determine upfront, such as classes and relations.

\paragraph{Star Patterns} Instead of estimating the completeness for a relation by looking only at that relation, we can look at facts involving \emph{other} relations.
For example, if someone has won a Nobel Prize, then we probably know their alma mater.
Formally, we consider ``star-shaped patterns'' of certain relations around the subject, and define the \emph{star oracle}, that predicts completeness if these patterns are all present:
\[\star_{r_1 \ldots r_n}(s, r)\defeq \forall i \in \{1, \ldots, n\} :  \exists o_i : r_i(s, o_i)\]

\paragraph{Class Information}
In some cases, the class to which an entity belongs can indicate completeness with respect to some relations.
For example, the instances of the class \emph{LivingPeople} should not have a death date. If we assume that the KB is correct, this implies that any instance of that class is complete with respect to that relation.
Formally, the \emph{class oracle} for a class expression $c$ on our KB $\mathcal{K}$ is:
\[\class_c(s, r)\defeq \type(s, c) \in \mathcal{K}\]

We conduct our study with two types of class expressions: plain class names such
as \emph{LivingPeople} and negated class expressions of the form $\hat{t} \land \neg t$ where $t$ is a subclass
of $\hat{t}$, like in $\mathit{Person} \land \neg\mathit{Adult}$.

\paragraph{Others} Many other completeness oracles can be envisaged. For example, we could extract information from the Web to find out whether we can find more objects; we could ask a crowd of users for more objects; we could compare two KBs to see if one contains more information than the other; or we 
could check against external sources.
In this paper, however, we limit ourselves to a single source, and leave other such approaches to future work.

\section{Learning Completeness}
\label{sec:learningcompleteness}
\subsection{Combining Oracles}\label{sec:combine}
Some completeness oracles cannot be used out-of-the-box. For example, to use the star oracle and the class oracle, we must try out a huge number of possible parameters: YAGO, e.g., has 200,000 classes. Furthermore, oracles may work best in combination: in some cases, the PCA may be the best oracle, while in others, the cardinality oracle may be better. Our goal is thus to \emph{generalize and learn} more complex completeness oracles from the simple ones that we presented.

Towards this goal, we assume that we already have a certain number of gold standard completeness assertions as \emph{training data}. We show in Section~\ref{sec:experiments} how to obtain such assertions from a crowd of users with good precision.
Based on these gold standard annotations, we can then learn combinations and parametrizations of the oracles.
To this end, we adapt the AMIE rule mining approach~\cite{amie,amie-plus}.

\subsection{AMIE}
\label{subsec:AMIE}
\paragraph{AMIE} AMIE~\cite{amie,amie-plus} is an inductive logic programming system that is particularly geared towards KBs.
The source code of AMIE is available online\footnote{\url{https://www.mpi-inf.mpg.de/departments/databases-and-information-systems/research/yago-naga/amie/}}.
Given a KB, AMIE finds Horn rules such as $\mathit{marriedTo}(X, Y) \wedge \mathit{livesIn}(X, Z) \Rightarrow \mathit{livesIn}(Y, Z)$. These rules do not hold in all cases, and therefore come with a \emph{confidence value}.

In AMIE, an \emph{atom} is a binary fact where at least one of the arguments is a variable~-- as in $\mathit{livesIn}(\mathit{Obama}, Y)$. We write the variables of atoms as capital letters.
A \emph{rule} is an expression of the form $\bm{B} \Rightarrow H$, where $\bm{B}$ is the \emph{body} (a conjunction of atoms $B_1 \land \dots \land B_n$), and $H$ is the \emph{head} (a single atom). 
The \emph{support} of a rule is the number of different instantiations of the head variables that satisfy all atoms of the rule in the KB. If $H=r(x,y)$, the support is defined by:
\[ supp(\bm{B} \Rightarrow r(x, y)) = \#(x, y) : \bm{B} \land r(x, y) \] 

\paragraph{Rule Mining} AMIE starts with rules with an empty body (i.e., rules
of the form ``${\Rightarrow r(X, Y)}$''), and refines them using a number of operators. Each of the operators takes a rule as input, and produces a set of refined rules as output, by adding one particular type of atom to the body of the rule:

\begin{itemize}
  \item \textbf{Add Dangling Atom:}
    A \emph{dangling atom} joins the rule on an existing variable and introduces a new variable in the other position.
  \item \textbf{Add Closing Atom:}
    A \emph{closing atom} is an atom that joins on two existing variables in the rule.
  \item \textbf{Add Instantiated Atom:}
    An \emph{instantiated atom} has one instantiated argument (a constant/entity) and joins with the rule on the other argument.
\end{itemize}
The operators always produce rules with less support than the original rule. AMIE applies them iteratively to find all rules above a given support threshold.

\subsection{Enhancing AMIE}
\label{subsec:enhancingAMIE}
\noindent Our goal is now to teach AMIE to learn rules such as
\[\mathit{moreThan}_1(X, \mathit{hasParent}) \Rightarrow \complete(X, \mathit{hasParent})\]
This rule says that if $X$ has more than one object for the relation \emph{has\-Parent},
then $X$ is probably complete on that relation. For this purpose, we assume that we have \emph{training data}, i.e. known assertions of the form $\complete(x, r)$ and $\incomplete(x, r)$. We show
in Section~\ref{sec:experiments} how to obtain such training data from the crowd. Then, all of the completeness oracles (Section~\ref{sec:estimatingcompleteness})
have to be translated into the AMIE framework. For this purpose, we define the following new types of atoms :
\begin{itemize}
  \item \textbf{\emph{complete}(\emph{x}, \emph{r}), \emph{incomplete}(\emph{x}, \emph{r}):} These assertions represent our training data. We add them to the KB.
  \item \textbf{\emph{isPopular}(\emph{x}):} The popularity oracle relies on an external measure $\pop$ of entity popularity. We considered three such measures: (i) number of facts for that entity, (ii) length of the article in the English Wikipedia, and (iii) number of ingoing links to the Wikipedia page. Manual inspection revealed that (i) correlated best with completeness. Thus, we add $\mathit{isPopular}(x)$ to the KB if $x$ is among the 5\% entities with the most facts in the KB.
  \item \textbf{\emph{hasNotChanged}(\emph{x}, \emph{r}):} Given an older version of the KB, we add the fact $\mathit{hasNotChanged}(x, r)$ to the new KB if $x$ has exactly the same $r$-objects in the new KB as in the old KB. In our experiments, we applied this to the YAGO KB, for which we used the oldest version (YAGO1) and the newest one (YAGO3).
  \item \textbf{\emph{notype}(\emph{x}, \emph{t}):}
The $\mathit{notype}(x, t)$
atom states that an entity is not an instance of class $t$.
Such atoms are always used in conjunction
with instantiated atoms of the form $type(x, \hat{t})$ where $\hat{t}$ is a super-class of $t$.
These types of atoms
allow us to integrate class expressions of the form $\hat{t} \land \neg t$ as defined for
the class oracle.
\item \textbf{\emph{lessThan}$_n$(\emph{x}, \emph{r}), \emph{moreThan}$_n$(\emph{x}, \emph{r}):}
An atom of the form $\mathit{lessThan}_n(x, r)$ with $n > 0$ is satisfied if $x$ has less than $n$ objects for relation $r$
in the KB. The \emph{moreThan}$_n$ atom is defined analogously. Such atoms allow AMIE to learn the cardinality
oracles that we introduced.
\end{itemize}
To use AMIE for the task of learning completeness, we also made some changes to
the system.
We let AMIE mine only rules with heads of the form $c(X, r)$, where $c$ is either $\complete$ or $\incomplete$, $r$ is a relation, and $X$ is a variable. 
We represent unary atoms $p(x)$ as $p(x, \true)$ since AMIE only supports binary atoms.
For performance reasons, we enable the ``Add Instantiated Atom''
operator only for $\mathit{isPopular}(x)$,
$\mathit{hasNotChanged}(x, r)$, $\mathit{type}(x, t)$ and $\mathit{notype}(x, t)$. 
Another problem is that AMIE's rule language enforces \emph{closed} Horn rules,
which are rules where each variable is \emph{closed}, i.e., 
it appears in at least two atoms in the rule.
We drop this constraint for variables in the body of rules, in order to allow for
rules with star patterns such as:
\[\mathit{wonPrize}(x, z) \wedge \mathit{politicianOf}(x, w) \Rightarrow \mathit{complete}(x, \mathit{citizenOf})\]
Still, we do not allow non-closed variables in the new kinds of atoms, e.g., $\mathit{isPopular}$ 
and $\mathit{hasNotChanged}$. We also forbid atoms with the relation $r$ in the body of the rules.
The last change is that we define five additional mining operators to capture the oracles that we defined:

\begin{itemize}
  \item \textbf{Add Type:}
    Given a rule $\bm{B} \Rightarrow c(X, r)$,
this operator adds an atom of the form $\type(X, t)$, where $t$ is the domain of $r$. The operator
is applied only if the rule does not yet contain a type atom.
\item \textbf{Specialize Type:}
  Given a rule $\type(X, t) \wedge \bm{B} \Rightarrow c(X, r)$,
  this operator yields a new rule $\type(X, t') \wedge \bm{B} \Rightarrow c(X, r)$
  where $t'$ is a subclass of~$t$.
\item \textbf{Add Negated Type:}
Given a rule $\type(X, t) \wedge \bm{B} \Rightarrow c(X, r)$,
this operator produces a new rule $\notype(X, t') \wedge \type(X, t) \wedge \bm{B} \Rightarrow c(X, r)$,
where $t'$ is a subclass of~$t$.
\item \textbf{Add Cardinality Constraint:}
  Given a rule $\bm{B} \Rightarrow c(X, r)$, this operator adds an atom of the form $\moreThan_0(X, r)$
or $\lessThan_k(X, r)$, where $k$ is
    the highest number of objects seen for any subject in the relation~$r$.
\item \textbf{Tighten Cardinality Constraint:}
  Given a rule $\lessThan_k(X, r) \wedge \bm{B} \Rightarrow c(X, r)$,
this operator replaces $k$ by the largest value $k'$ (with $k' < k$) that decreases the support
of the original rule. Likewise, given a rule $\moreThan_k(X, r) \wedge \bm{B} \Rightarrow c(X, r)$,
we replace $k$ by the smallest value $k'$ ($> k$) that decreases the support. For example,
given the rule $\moreThan_0(X, \mathit{hasParent}) \Rightarrow \complete(X, \mathit{hasParent})$,
the operator will replace 0 by~1.
\end{itemize}

\paragraph{Learning} With our supplementary atoms and new mining operators, and up
to the changes that we described, the actual learning of completeness rules works exactly as the mining of normal rules in~\cite{amie,amie-plus}.
We exemplify this by showing how AMIE mines the following rules:
\begin{enumerate}
 \item $\notype(X, \mathit{Adult}) \wedge \type(X, \mathit{Person})
   \Rightarrow \complete(X, \mathit{hasChild})$
 \item $\mathit{lessThan_1}(X, \mathit{isCitizenOf}) \Rightarrow \mathit{incomplete}(X, \mathit{isCitizenOf})$
\end{enumerate}
The first rule says that if a person is not an adult, then the KB is complete for the children of
that person (most likely zero). To mine this rule, AMIE starts with the simple rule ``$\Rightarrow \complete(X, \mathit{hasChild})$'' and applies
all the mining operators described in Sections~\ref{subsec:AMIE} and \ref{subsec:enhancingAMIE}.
Among the different new rules generated by this step, the ``Add Type'' operator
produces the rule
$\mathit{type}(X, \mathit{Person}) \Rightarrow \complete(X, \mathit{hasChild})$.
In the next step, the operator ``Add Negated Type''
produces new rules of the form $\notype(X, t) \wedge \type(X, \mathit{Person}) \Rightarrow \complete(X, \hasChild)$,
where $t$ is a subclass of $\mathit{Person}$. In particular, for $t=\mathit{Adult}$, we obtain our example rule.

The second rule states that if a person has less than one citizenship, then the KB is incomplete in the citizenship relation for that person. 
AMIE starts with the rule $\Rightarrow \mathit{incomplete}(X, \mathit{isCitizenOf})$, and applies the 
``Add Cardinality Constraint''. Assuming that in the KB nobody has more than 3 nationalities, 
the operator produces
the rule $\mathit{lessThan_3}(X, \mathit{isCitizenOf}) \Rightarrow \mathit{incomplete}(X, \mathit{isCitizenOf})$.
This rule has support~$s$.
In a later step, AMIE tries to apply the `Tighten Cardinality Constraint'' operator.
The operator searches for the largest $k < 3$ such that the support of the new rule drops. 
If the number of incomplete people with less than 2 nationalities is smaller than $s$, the system will chose $k= 2$ 
and the rule becomes $\mathit{lessThan_2}(X, \mathit{isCitizenOf}) \Rightarrow \mathit{incomplete}(X, \mathit{isCitizenOf})$. 
Using again the ``Tighten Cardinality Constraint'' operator on the new rule
produces
$\mathit{lessThan_1}(X, \mathit{isCitizenOf}) \Rightarrow \mathit{incomplete}(X, \mathit{isCitizenOf})$.
We remark that depending on the data distribution, AMIE may need a single call to the 
``Tighten Cardinality Constraint'' to produce the target rule, i.e., we may skip the intermediate step where $k=2$.

\paragraph{AMIE as completeness oracle}
AMIE will learn rules that predict completeness as well as rules that predict incompleteness.
For the first type of rules, AMIE uses the $\complete(x, r)$ atoms of the training data as examples,
and the $\incomplete(x, r)$ atoms as counter-examples. For the second type of rules, the roles are reversed.
This implies that confidence for completeness and incompleteness rules follows the formula:
\[\textit{conf}(\bm{B} \Rightarrow c(X, r)) = \frac{supp(\bm{B} \Rightarrow c(X,
r))}{supp(\bm{B} \Rightarrow c(X, r)) + supp(\bm{B} \Rightarrow \neg c(X, r)) }
\]
where $c \in \{ \mathit{complete}, \mathit{incomplete} \}$.

Once the rules have been learned, we can define a new completeness oracle, the \emph{AMIE oracle}. For a given entity~$e$ and a relation~$r$, the AMIE oracle checks whether any of the learnt rules predicts $\complete(e, r)$.
If so, and if there is no rule with higher or equal confidence that predicts $\incomplete(e, r)$,
the oracle returns \emph{true}.
If there is a rule with equal confidence that predicts $\incomplete(e, r)$, the oracle returns \emph{true} if the support of the completeness rule is higher. In all other cases, the oracle returns \emph{false}.

By restricting AMIE to only star atoms or only class atoms, we can obtain a \emph{Star oracle} and a \emph{Class oracle}, respectively, analogously to the AMIE oracle.

\section{Experiments}
\label{sec:experiments}
\begin{table*}[t]
\begin{center}
  \begin{tabular}{l}
   \toprule
    \emph{dateOfDeath(X, Y)} $\wedge$ \emph{lessThan$_1$(X, placeOfDeath)}  $\Rightarrow$ \emph{incomplete(X, placeOfDeath)}\\
\emph{IMDbId(X, Y)} $\wedge$ \emph{producer(X, Z)}  $\Rightarrow$ \emph{complete(X, director)}\\
\emph{notype(X, Adult)} $\wedge$ \emph{type(X, Person)} $\Rightarrow$ \emph{complete(X, hasChild)}\\
\emph{lessThan$_2$(X, hasParent)} $\Rightarrow$ \emph{incomplete(X, hasParent)}\\
   \bottomrule
  \end{tabular}
  \caption{Example rules that AMIE learned (2 on Wikidata, 2 on YAGO)}
  \label{rules}
  \end{center}
\end{table*}

\subsection{Setup}

\paragraph{Knowledge bases} Our goal is to measure the precision and recall of the completeness oracles on real data. 
We conducted our study on two KBs: YAGO3, released in September 2015, and a dump of Wikidata from December 2015.
For both datasets, we used the facts between entities, the facts with literal object values (except for the relation \emph{rdfs:label}) 
and the instance information. These choices leave us with a KB of 89M facts (78M type statements) for YAGO,
and a KB of 15M facts (3.6M type statements) for Wikidata.
We studied completeness on a set of relations covering a large variety of cases,
and including people, movies, and locations:
\begin{itemize}
  \item For one type of relations, basically every entity of the domain has to have exactly one object:
\emph{hasGender}, \emph{wasBornIn} in YAGO; \emph{sex\_or\_gender} (P21), \emph{mother} (P25), \emph{father} (P22), \emph{place\_of\_birth} (P19) in Wikidata.
  \item For some other relations, entities do not need to have an object, but can have at most one: \emph{diedIn} in YAGO; \emph{place\_of\_death} (P20) in Wikidata.
  \item Some other relations usually have one object, but can have more: \emph{isCitizenOf} and \emph{director}(\emph{Movie}, \emph{Person}) in YAGO;
\emph{country\_of\_citizenship} (P27) and \emph{director} (P57) in Wikidata.
\item In the most general case, a subject can have zero, one, or several objects: \emph{hasChild}, 
\emph{graduatedFrom}, \emph{isConnectedTo}(\emph{Airport}, \emph{Airport}), and \emph{isMarriedTo}\footnote{Despite the name, this relation captures also past spouses.} in YAGO; \emph{child} (P40), \emph{alma\_mater}\footnote{We use the same semantics as in YAGO: places a person graduated from.} (P69), \emph{brother}, and \emph{spouse} (P26) in Wikidata.
    \item One relation has to have 2 objects: \emph{hasParent}\footnote{This is how we call the inverse of \emph{hasChild} in YAGO.} in YAGO.
      \end{itemize}

\paragraph{Ground Truth} In order to evaluate our completeness oracles, we need
a set of completeness assertions and incompleteness assertions as a gold
standard. For some relations, we could generate this gold standard
automatically.
Namely, for the relations where every subject has to have exactly one object,
we have $\complete(s, r)$ iff $\exists o: r(s, o)$. For the relations where
every subject must have at least one object,
we can directly label as incomplete all subjects without a value. For the relations with at most one object,
all subjects with one object are considered complete.
For the relation \emph{isConnectedTo}, we used the
OpenFlights\footnote{\url{http://openflights.org/data.html}} dataset as ground
truth, which we assumed to be complete for all airports in this dataset
(identified by their IATA code). However, due to data inaccuracies,
for some airports YAGO knew more flights than OpenFlights:
we discarded these airports.

\paragraph{Crowdsourcing} For the remaining relations, we used \emph{crowdsourcing} to obtain ground truth data.
Given an entity, a relation, and the objects known in the KB, we asked crowd workers whether they could find any additional objects on the Web. If they could, we labelled the entity-relation pair as incomplete, otherwise as complete. To make the task well-defined and manageable, we asked workers to look only at a set of given Web pages. We manually defined queries for each relation (e.g., ``$x$ died'' for $\mathit{diedIn}(x, y)$ or ``$x$ child'' for $\mathit{hasChild}(x, y)$), and then gave workers the first 4 URLs retrieved using the Bing search API. We used the Crowdflower platform\footnote{\url{https://www.crowdflower.com}} for crowdsourcing, and paid 1 cent per answer. For every relation, 
we annotated 200 random entities.
For each entity, we collected 3 opinions.

\paragraph{Quality Control} To monitor quality, we manually generated \mbox{20--29} test questions for each relation. Annotators had to pass a qualification test of 10 questions with at least 80\% correct answers; further, the remaining test questions were mixed with the data, and annotators had to maintain 80\% correctness while working. About a quarter of annotators failed at the initial tests, and about 5\% fell below the correctness threshold while working.
Their answers were discarded. Furthermore, we used only the
annotations where all 3 answers were unanimous. These make up 55\% of the annotations.

\paragraph{Sampling} 
In our experiments with AMIE, we use 80\%
of our gold standard for training, and the rest for testing.
This gold standard was produced by randomly picking 200
entities in the domain of the studied relations. We call this
sample uniform.
The uniform sample is not always useful. For example, only 1\% of people
have a citizenship in YAGO. Thus, in a sample of 200 people, we may expect a citizenship for only 2 of them, which is not enough to learn a rule. Therefore,
for relations where less
than 10\% of the subjects have an object,
we construct a \emph{biased sample} instead.
Rather than choosing 200 entities randomly, we choose 100 entities randomly among those that have an object,
and 100 among those that do not. In our experiments, we mark the relations where we used the biased sample. 
For the calculation of precision and recall, we carried out a de-biasing step. 
This means that the values we report reflect the true population of entities in the KBs, and not the biased population.

\subsection{Basic Completeness Oracles}

\paragraph{Experiment}  Our completeness oracles from Section~\ref{sec:estimatingcompleteness} try to guess whether  a pair of a subject and a relation is complete. We considered the subject--relation pairs where we had a gold standard, and computed precision and recall values as described in Section \ref{sec:preliminaries}. Table~\ref{tab:all-pr-yago} shows the results for the oracles for YAGO3, and Table~\ref{tab:all-pr-wikidata} for Wikidata. Table~\ref{tab:all-f1-yago} and Table~\ref{tab:all-f1-wikidata} show the corresponding F1 measures.

\paragraph{Cardinality Oracles} The first column in the tables shows the CWA.
It trivially achieves a recall of 100\%: for all pairs that are complete in reality, it makes a correct prediction. However, its precision is lower. This precision value corresponds to the actual completeness of the KB with respect to the real world. We see, e.g., that YAGO is complete for the death place for 44\% of the people. This means that these people are either alive, or dead with a known death place in YAGO.
We also observe that Wikidata is generally more complete than YAGO.

The next oracle is the PCA. It achieves 100\% precision for all functional relations: if a subject has an object, the PCA rightly assumes that the subject is complete. For quasi-functions, such as \emph{is\-Citizen\-Of}, the PCA still performs decently, failing only for people with several nationalities. The PCA has a recall of 100\% for relations that are mandatory (such as \emph{has\-Gender}): whenever this relation is complete in the gold standard, the PCA indeed predicts it. For the other relations, the PCA has a much lower precision and recall.

The $card_2$ oracle has a much lower recall. We could not compute it for relations where the  sample did not contain any entity with sufficiently many objects.
This oracle basically makes sense only for the \emph{has\-Parent} relation, where it performs perfectly. As $\card_3$ behaves worse that $card_2$ on both datasets, we omitted it for space reasons.

\paragraph{Popularity Oracle} The fourth column shows the popularity oracle. The oracle was not computed for \emph{isConnectedTo} due to noise in the data. The popularity oracle generally has a low recall, because there are not many popular entities. Its precision is generally good, indicating that popular entities (those that have many facts in general) are indeed more complete than unpopular ones. However, even popular entities are incomplete for parents and citizenship in YAGO, and for parents in Wikidata.

\paragraph{No-Change Oracle} The next column shows the no-change oracle on YAGO, for those relations that exist in both YAGO1 and YAGO3. It has a very low recall, indicating that most entities did indeed change their objects over time (they most likely gained more objects). The precision is decent, but not extraordinary.

\begin{table*}[t]
\centering
  \footnotesize \renewcommand{\tabcolsep}{-1pt}
\begin{tabularx}{\linewidth}{X
  *{8}{@{\hspace{-.1cm}}*{8}{P{.85cm}}}
  }
  \toprule
{\bf Relation} &
  \multicolumn{2}{c}{\bf CWA} &
  \multicolumn{2}{c}{\bf PCA} &
  \multicolumn{2}{c}{\bf $\mathbf{\text{\emph{\bfseries card}}_2}$} &
  \multicolumn{2}{c}{\bf Popularity~} &  
  \multicolumn{2}{c}{\bf ~No-change} &  
  \multicolumn{2}{c}{\bf Star} &
  \multicolumn{2}{c}{\bf Class} &
  \multicolumn{2}{c}{\bf AMIE} \\
& \multicolumn{1}{c}{\bf Pr} & \multicolumn{1}{c}{\bf Rec} 
& \multicolumn{1}{c}{\bf Pr} & \multicolumn{1}{c}{\bf Rec} 
& \multicolumn{1}{c}{\bf Pr} & \multicolumn{1}{c}{\bf Rec}
& \multicolumn{1}{c}{\bf Pr} & \multicolumn{1}{c}{\bf Rec}
& \multicolumn{1}{c}{\bf Pr} & \multicolumn{1}{c}{\bf Rec}
& \multicolumn{1}{c}{\bf Pr} & \multicolumn{1}{c}{\bf Rec}
& \multicolumn{1}{c}{\bf Pr} & \multicolumn{1}{c}{\bf Rec}
& \multicolumn{1}{c}{\bf Pr} & \multicolumn{1}{c}{\bf Rec}
\\
\midrule
diedIn &  43\% & 100\% &  100\% & 13\% &  --- & --- &  97\% & 2\% & 74\% & 8\% &  100\% & 33\% &  100\% & 97\% &  96\% & 96\% \\
directed &  25\% & 100\% &  93\% & 100\% &  72\% & 11\% &  91\% & 3\% & 90\% & 59\% &  0\% & 0\% &  0\% & 0\% &  100\% & 100\% \\
graduatedFrom &  80\% & 100\% &  70\% & 2\% &  79\% & 1\% &  89\% & 1\% & 82\% & 6\% &  84\% & 94\% &  85\% & 100\% &  77\% & 100\% \\
hasChild &  55\% & 100\% &  36\% & 1\% &  41\% & 0\% &  78\% & 1\% & 70\% & 7\% &  83\% & 26\% &  63\% & 100\% &  65\% & 100\% \\
hasGender &  64\% & 100\% &  100\% & 100\% &  --- & --- &  98\% & 1\% & --- & --- &  92\% & 81\% &  91\% & 100\% &  100\% & 100\% \\
hasParent* &  0\% & 100\% &  37\% & 100\% &  100\% & 100\% &  --- & --- & --- & --- &  0\% & 0\% &  0\% & 0\% &  100\% & 100\% \\
isCitizenOf* &  2\% & 100\% &  97\% & 100\% &  93\% & 6\% &  2\% & 1\% & 2\% & 7\% &  6\% & 33\% &  2\% & 53\% &  100\% & 100\% \\
isConnectedTo &  77\% & 100\% &  67\% & 23\% &  60\% & 12\% &  --- & --- & --- & --- &  77\% & 62\% &  79\% & 100\% &  81\% & 100\% \\
isMarriedTo* &  38\% & 100\% &  84\% & 4\% &  92\% & 0\% &  66\% & 1\% & 51\% & 7\% &  25\% & 74\% &  40\% & 100\% &  29\% & 100\% \\
wasBornIn &  16\% & 100\% &  100\% & 100\% &  --- & --- &  73\% & 3\% & 33\% & 5\% &  0\% & 0\% &  0\% & 0\% &  100\% & 100\% \\
\bottomrule
\end{tabularx}

  \caption{Precision and recall of all completeness oracles on YAGO3. Relations with a biased sample are marked with *.}
  \label{tab:all-pr-yago}
\end{table*}

\begin{table*}[t]
\centering
  \begin{tabularx}{\linewidth}{lcccccccc}
  \toprule
  {\bf Relation} & {\bf CWA} 	& {\bf PCA} 	 &
  {\bf $\mathbf{\text{\emph{\bfseries card}}_2}$}
  & {\bf Pop.} 	& {\bf No-ch.} & {\bf Star} 	& {\bf Class} 	& {\bf AMIE}\\
\midrule
diedIn &  60\% &  22\% &  --- &  4\% & 15\% &  50\% &  \textbf{99\%} &  96\% \\
directed &  40\% &  96\% &  19\% &  7\% & 71\% &  0\% &  0\% &  \textbf{100\%} \\
graduatedFrom &  89\% &  4\% &  2\% &  2\% & 10\% &  89\% &  \textbf{92\%} &  87\% \\
  hasChild &  71\% &  1\% &  1\% &  2\% & 13\% &  40\% &  \textbf{78\%} &  \textbf{78\%} \\
hasGender &  78\% &  \textbf{100\%} &  --- &  2\% & --- &  86\% &  95\% &  \textbf{100\%} \\
hasParent* &  1\% &  54\% &  \textbf{100\%} &  --- & --- &  0\% &  0\% & \textbf{100\%} \\
isCitizenOf*
  &  4\% &  98\% &  11\% &  1\% & 4\% &  10\% &  5\% &  \textbf{100\%} \\
isConnectedTo &  87\% &  34\% &  19\% &  --- & --- &  68\% &  88\% &  \textbf{89\%} \\
isMarriedTo* &  55\% &  7\% &  0\% &  3\% & 12\% &  37\% &  \textbf{57\%} &  46\% \\
wasBornIn &  28\% &  \textbf{100\%} &  --- &  5\% & 8\% &  0\% &  0\% &  \textbf{100\%} \\
\bottomrule
\end{tabularx}

  \caption{F1 measure of all completeness oracles on YAGO3. Relations with a biased sample are marked with *.}
  \label{tab:all-f1-yago}
\end{table*}

\begin{table*}[t]
\centering
  \footnotesize \renewcommand{\tabcolsep}{-1pt}
\begin{tabularx}{\linewidth}{X
  *{7}{@{\hspace{-.1cm}}*{7}{P{.9cm}}}
  }
  \toprule
{\bf Relation} &
  \multicolumn{2}{c}{\bf CWA} &
  \multicolumn{2}{c}{\bf PCA} &
  \multicolumn{2}{c}{\bf $\mathbf{\text{\emph{\bfseries card}}_2}$} &
  \multicolumn{2}{c}{\bf Popularity} &  
  \multicolumn{2}{c}{\bf Star} &
  \multicolumn{2}{c}{\bf Class} &
  \multicolumn{2}{c}{\bf AMIE} \\
& \multicolumn{1}{c}{\bf Pr} & \multicolumn{1}{c}{\bf Rec} 
& \multicolumn{1}{c}{\bf Pr} & \multicolumn{1}{c}{\bf Rec} 
& \multicolumn{1}{c}{\bf Pr} & \multicolumn{1}{c}{\bf Rec}
& \multicolumn{1}{c}{\bf Pr} & \multicolumn{1}{c}{\bf Rec}
& \multicolumn{1}{c}{\bf Pr} & \multicolumn{1}{c}{\bf Rec}
& \multicolumn{1}{c}{\bf Pr} & \multicolumn{1}{c}{\bf Rec}
& \multicolumn{1}{c}{\bf Pr} & \multicolumn{1}{c}{\bf Rec}
\\
\midrule
alma\_mater &  82\% & 100\% &  80\% & 8\% &  95\% & 2\% &  57\% & 1\%  &  76\% & 100\% &  76\% & 100\% &  76\% & 100\% \\
brother &  86\% & 100\% &  57\% & 0\% &  --- & --- &  61\% & 1\%  &  92\% & 96\% &  92\% & 100\% &  92\% & 100\% \\
child &  54\% & 100\% &  15\% & 1\% &  --- & --- &  25\% & 0\%  &  73\% & 86\% &  58\% & 95\% &  79\% & 68\% \\
country\_of\_citizenship* &  27\% & 100\% &  95\% & 100\% &  100\% & 5\% &  38\% & 1\%  &  0\% & 0\% &  0\% & 0\% &  96\% & 100\% \\
director &  68\% & 100\% &  100\% & 100\% &  --- & --- &  95\% & 1\%  &  89\% & 100\% &  85\% & 94\% &  100\% & 100\% \\
father* &  3\% & 100\% &  100\% & 100\% &  100\% & 3\% &  16\% & 6\%  &  100\% & 80\% &  4\% & 82\% &  100\% & 100\% \\
mother* &  1\% & 100\% &  100\% & 100\% &  100\% & 1\% &  12\% & 9\%  &  52\% & 96\% &  2\% & 86\% &  100\% & 100\% \\
place\_of\_birth &  36\% & 100\% &  100\% & 100\% &  100\% & 4\% &  90\% & 3\%  &  86\% & 41\% &  0\% & 0\% &  100\% & 100\% \\
place\_of\_death &  81\% & 100\% &  100\% & 21\% &  100\% & 1\% &  97\% & 1\%  &  77\% & 87\% &  77\% & 87\% &  93\% & 100\% \\
sex\_or\_gender &  69\% & 100\% &  100\% & 100\% &  100\% & 3\% &  96\% & 1\%  &  87\% & 98\% &  85\% & 97\% &  100\% & 100\% \\
spouse* &  40\% & 100\% &  88\% & 4\% &  --- & --- &  29\% & 1\%  &  38\% & 99\% &  37\% & 99\% &  38\% & 100\% \\
\bottomrule
\end{tabularx}

  \caption{Precision and recall of all completeness oracles on Wikidata. Relations with a biased sample are marked with *.}
  \label{tab:all-pr-wikidata}
\end{table*}

\begin{table*}[t]
\centering
  \begin{tabularx}{\linewidth}{Xccccccc}
  \toprule
{\bf Relation} & {\bf CWA} 	& {\bf PCA} 	
  & {\bf $\mathbf{\text{\emph{\bfseries card}}_2}$}
  & {\bf Pop.}  & {\bf Star} 	& {\bf Class} 	& {\bf AMIE}\\
\midrule
alma\_mater &  \textbf{90\%} &  14\% &  5\% &  1\%  &  87\% &  87\% &  87\% \\
brother &  93\% &  1\% &  --- &  1\%  &  94\% &  \textbf{96\%} &  \textbf{96\%} \\
child &  70\% &  1\% &  --- &  1\%  &  \textbf{79\%} &  72\% &  73\% \\
country\_of\_citizenship* &  42\% &  97\% &  10\% &  3\%  &  0\% &  0\% &  \textbf{98\%} \\
director &  81\% &  \textbf{100\%} &  --- &  3\%  &  94\% &  89\% &  \textbf{100\%} \\
father* &  5\% &  \textbf{100\%} &  6\% &  9\%  &  89\% &  8\% &  \textbf{100\%} \\
mother* &  3\% &  \textbf{100\%} &  3\% &  10\%  &  67\%* &  5\% &  \textbf{100\%} \\
place\_of\_birth &  53\% &  \textbf{100\%} &  7\% &  5\%  &  55\% &  0\% &  \textbf{100\%} \\
place\_of\_death &  89\% &  35\% &  1\% &  2\%  &  81\% &  81\% &  \textbf{96\%} \\
sex\_or\_gender &  81\% &  \textbf{100\%} &  6\% &  3\%  &  92\% &  91\% &  \textbf{100\%} \\
spouse* &  \textbf{57\%} &  7\% &  --- &  1\%  &  54\% &  54\% &  55\% \\
\bottomrule
\end{tabularx}

  \caption{F1 measure of all completeness oracles on Wikidata. Relations with a biased sample are marked with *.}
  \label{tab:all-f1-wikidata}
\end{table*}

\subsection{Learned Completeness Oracles}\label{sec:learning-experiment}

\paragraph{Learning} We took 80\% of our gold standard to train our modified AMIE approach
(Section~\ref{sec:learningcompleteness}) with 4-fold cross-validation. 
The training phase measures the performance of AMIE at different configurations, i.e., different values for the
support and confidence thresholds.
We tested values for support in the range from 10 to 100 entities (in steps of~10), while confidence was tested on 
values from 0.3 to 1.0 (in steps of 0.1).
We report the best configuration in terms of F1 measure for each relation, and use it to measure performance in the testing set 
(the remaining 20\% of the gold standard).
Training took 44 hours on YAGO,
and 4 hours in Wikidata. This difference is mainly due to the much larger type hierarchy in YAGO
(78M type assertions as opposed to 3.6M in Wikidata). Table~\ref{rules} shows some of the rules that AMIE learned.
The first rule says that a person who has a date of death, but no place of death, is incomplete for the place of death.
In the second rule, the IMDb id acts as a substitute for the type \emph{movie}, which is not always consistently used in Wikidata. Thus, the rule basically says that if a movie has a producer, 
then it is most likely complete on the director. Many of our rules are specific to our dataset. Others (such as the first) may apply to different datasets. 
We leave the study of cross-dataset rules for future work, and concentrate on each individual dataset here.

\paragraph{Results} After the rules have been learned, making the actual oracle predictions on the gold standard takes only seconds. We evaluated these predictions against the remaining 20\% of our gold standard, and report the precision, recall, and F1 values in the three last columns of Tables~\ref{tab:all-pr-yago} and \ref{tab:all-f1-yago} for YAGO,
and in Tables~\ref{tab:all-pr-wikidata} and \ref{tab:all-f1-wikidata} for Wikidata.

For the star oracle, we used a star size of $n=1$ for YAGO and $n=3$ for Wikidata. We observe that this oracle
can improve the F1 value for the \emph{is\-Married\-To} relation. The class oracle, likewise, performs well for certain relations. In particular, the oracle learned that the YAGO class \emph{Living\-People} 
means that the \emph{died\-In} relation must be complete,
boosting F1 from 60\% to 99\%. This shows that parametrized oracles can be useful.

In general, the oracles complement each other. Only the complete AMIE approach
can nearly always perform best. This is because AMIE learns the strengths of the individual oracles, and combines them as is most useful. For functional relations,
AMIE learned a rule that mimics the PCA, predicting completeness for a subject
whenever one object is present: $\moreThan_0(X, r) \Rightarrow \complete(X, r)$. For \emph{died\-In}, AMIE learned a rule that mimics the Class oracle: $type(X, \mathit{LivingPeople}) \Rightarrow complete(X, \mathit{diedIn})$.
In this way, our oracle achieves an F1-measure of over 90\% for more than half of the relations -- on both YAGO and Wikidata.
When such relation-specific rules are not available, AMIE learns the CWA. This is the case 
for difficult relations
such as \emph{brother}, \emph{graduated\-From} or \emph{is\-Connected\-To}. In particular, AMIE learns the CWA 
in rules of the form 
\[\type(X, \mathit{domain}(r)) \Rightarrow \mathit{complete}(X, r)\]

All in all, our results show that it is indeed possible 
to predict completeness with very good precision and recall for a large number of relations. 
We can predict whether people are alive, whether they graduated, or whether they have siblings -- all by just looking at the incomplete KB. The \emph{has\-Child} and \emph{married\-To} relations are the only ones where our oracles perform less well. However, guessing whether someone is married, or whether someone has children, is close to impossible even for a human.

\section{Application}
\label{sec:usingcompleteness}
Having studied the experimental performance of our approach, we now show how the
completeness assertions that we generate can prove useful in applications. We
focus on \emph{fact prediction}, which we first define.

\paragraph{Goal} 
Rule mining is generally used to find arbitrary rules in a KB, not just completeness rules.
We can use these rules to perform \emph{fact prediction}, i.e., predict which person lives where, or which city is located in which country~\cite{amie,amie-plus}. We can compare the predictions to the real world and thus measure the precision of the approach.

We will show how the precision of fact prediction can be improved by completeness assertions. For this purpose, we use the standard AMIE approach to make fact predictions, but we use the completeness assertions to filter out some of them: we filter out predicted facts $r(s, o)$ whenever  $\complete(s, r)$ holds. For example, if fact prediction says that a person has a parent, but the KB already knows two parents, then we discard the prediction.

\begin{figure}
\begin{center}
      \includegraphics[width=\textwidth]{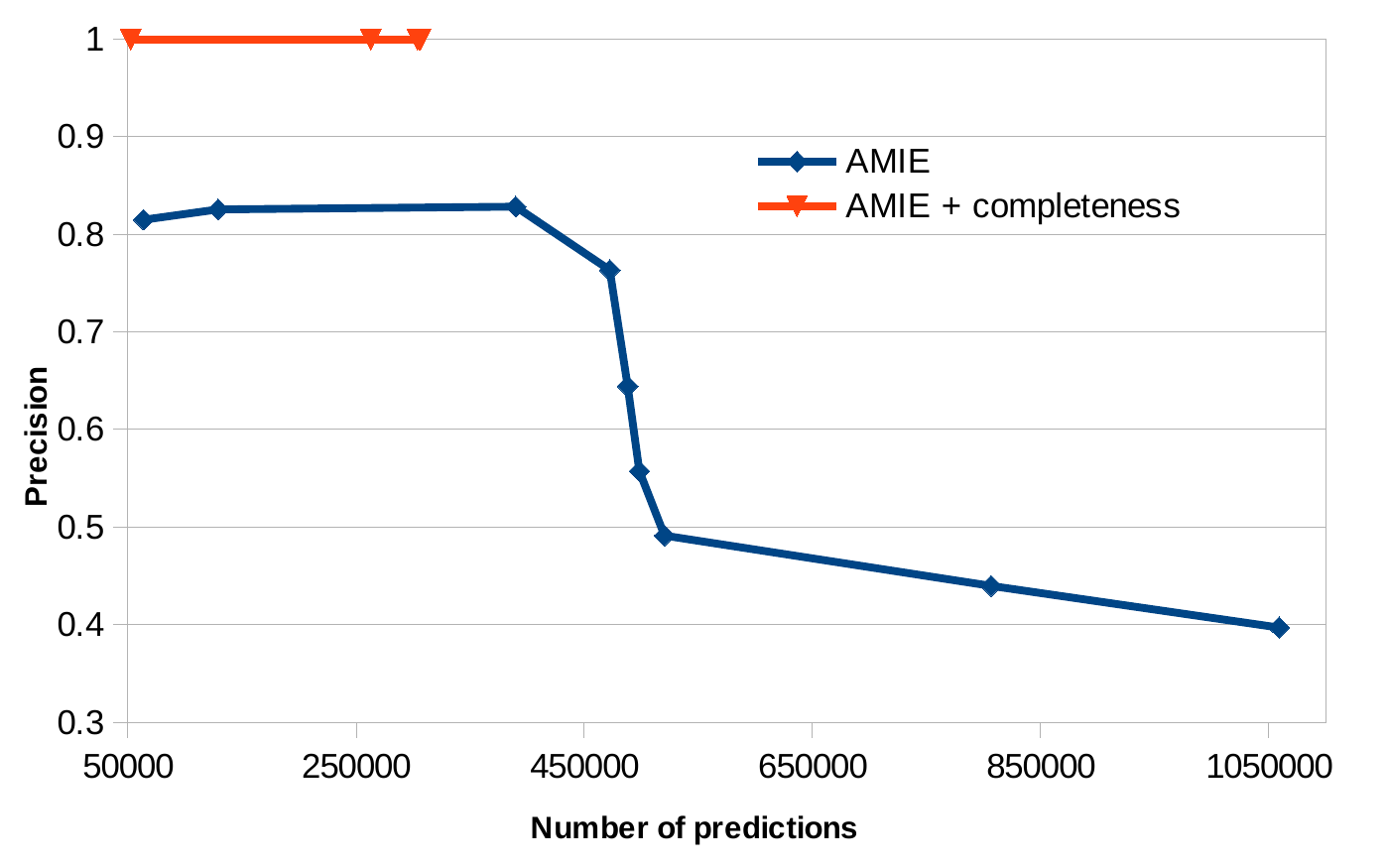}
      \caption{Precision of fact prediction}
      \label{fig:precisionYAGO}
\end{center}
\end{figure}

\paragraph{Setup} We followed the experimental setup from~\cite{amie-plus} and ran the standard AMIE system on YAGO3, using the obtained rules to predict new facts. Each rule (and thus each prediction) comes with a confidence score. We grouped the predictions in buckets by confidence score, as in~\cite{amie-plus}. For each bucket, we resorted to crowd workers to evaluate the precision of the predictions on a sample of 100 facts. The lower line in Figure~\ref{fig:precisionYAGO} shows the number of predictions versus the cumulative precision estimated on the samples. Each data point corresponds to a bucket of predictions, i.e., the first point on the left corresponds to the predictions with confidence score between 0.9 and 1, the second point to those with confidence between 0.8 and 0.9, etc. In the second phase of the experiment, we used completeness assertions to filter out predictions. We produced completeness assertions as in Section~\ref{sec:learning-experiment}, by training AMIE with cross-validation on our entire set of gold standard completeness assertions. The upper line in Figure~\ref{fig:precisionYAGO} shows the cumulative precision and number of predictions for each bucket after filtering.

\paragraph{Results} As we can observe, the filtering could successfully prune all wrong predictions. The remaining predictions have a precision of 100\%. This high precision has to be taken with a grain of salt: the remaining predictions are mainly about citizenship, which is guessed from the place of residence or place of birth. The completeness assertions filter out any predictions that try to assign a second citizenship to a person, and thus drastically increase the precision. However, there are also a few other relations among the predictions. These are, e.g., the death place, or the alma mater (guessed from the workplace of the academic advisor).

This precision comes at a price. In total, AMIE made 1.05M predictions. Of these, 400K were correct. From these, the filtering incorrectly removed 110K. Thus, the filtering removes roughly 25\% of the correct predictions as a side-effect. Still, we believe that our experiments make the case that completeness assertions can significantly improve the performance of fact prediction.

\section{Conclusion}
\label{sec:conclusion}
To the best of our knowledge, our work is the first systematic study of the problem of completeness in knowledge bases. Completeness is an important dimension of quality, which is orthogonal to the dimension of correctness, and which has so far received less attention. In our paper, we have defined and analyzed a range of simple and parametrized completeness oracles. We have also shown how to combine these oracles into more complex oracles by rule mining. Our experiments on YAGO and Wikidata prove that completeness can indeed be predicted with high precision for many relations. These completeness estimations can then be used to improve fact prediction to 100\% precision in specific cases.

We hope that our work can lead to new research avenues, aiming to design knowledge bases that are not only highly accurate, but also highly complete.
The experimental results of this paper are available at \url{http://luisgalarraga.de/completeness-in-kbs}.

\paragraph{Acknowledgment} This work has been partially supported by the project ``MAGIC'', funded by the Province of Bozen-Bolzano, and ``TQTK'', funded by the Free University of Bozen-Bolzano.

\bibliographystyle{plain}
\bibliography{references}

\end{document}